\documentclass[prd,showpacs,showkeys,floatfix,amsmath,amssymb,floatfix,english]{revtex4}

\usepackage{amsfonts}
\usepackage{subfig}
\usepackage{dcolumn}
\usepackage{natbib}
\usepackage{bm}
\usepackage{booktabs}
\usepackage{slashed}
\usepackage[T1]{fontenc}
\usepackage[utf8]{inputenc}
\usepackage{babel}
\usepackage{units}
\usepackage{amsmath}
\usepackage{esint}
\usepackage{multirow}
\usepackage{graphicx}
\usepackage[normalem]{ulem}
\usepackage[colorlinks, citecolor=blue,anchorcolor=red,menucolor=red, linkcolor=red,filecolor=red,runcolor=red,urlcolor=blue,frenchlinks=red]{hyperref}
\usepackage{amsmath}

\allowdisplaybreaks[4]

\usepackage{ulem}
\usepackage{color}

\makeatletter

\newcommand{\Rmnum}[1]{\expandafter\@slowromancap\romannumeral #1@}

\def\hlinewd#1{%
  \noalign{\ifnum0=`}\fi\hrule \@height #1 \futurelet
   \reserved@a\@xhline}
\makeatother

\begin{document}

\title{Exotic fully-heavy $Q\bar QQ\bar Q$ tetraquark states in $\mathbf{8}_{[Q\bar{Q}]}\otimes \mathbf{8}_{[Q\bar{Q}]}$ color configuration}

\author{Qi-Nan Wang}
\author{Zi-Yan Yang}

\author{Wei Chen}
\email{chenwei29@mail.sysu.edu.cn}
\affiliation{School of Physics, Sun Yat-Sen University, Guangzhou 510275, China}

\begin{abstract}
We have systematically calculated the mass spectra for S-wave and P-wave fully-charm $c\bar{c}c\bar{c}$ and fully-bottom $b\bar{b}b\bar{b}$ tetraquark states in the $\mathbf{8}_{[Q\bar{Q}]}\otimes \mathbf{8}_{[Q\bar{Q}]}$ color configuration, by using the moment QCD sum rule method. The masses for the fully-charm $c\bar cc\bar c$ tetraquark states are predicted about $6.3-6.5$ GeV for S-wave channels and $7.0-7.2$ GeV for P-wave channels. 
These results suggest the possibility that there are some $\mathbf{8}_{[c\bar{c}]}\otimes \mathbf{8}_{[c\bar{c}]}$ components in LHCb's di-$J/\psi$ structures. For the fully-bottom $b\bar{b}b\bar{b}$ system, their masses are calculated around 18.2 GeV for S-wave tetraquark states while 18.4-18.8 GeV for P-wave ones, which are below the $\eta_b\eta_b$ and $\Upsilon(1S)\Upsilon(1S)$ two-meson decay thresholds.
\end{abstract}


\maketitle

\section{Introduction}
The existence of multiquark states was first suggested by Gell-Mann and Zweig at the birth of quark model~\cite{Gell-Mann:1964ewy,ZweigSU(3)}. 
Since 2003, plenty of charmoniumlike exotic states and $P_{c}$ states have been observed~\cite{Belle:2003nnu,BESIII:2013mhi,BESIII:2013ris,BaBar:2005hhc,BaBar:2006ait,Belle:2011aa,Belle:2007umv,LHCb:2015yax,LHCb:2019kea,LHCb:2020jpq} , many of which are unexpected and can not be fitted into the conventional quark model. To understand the nature of these new resonances, many exotic hadron configurations have 
been proposed such as hadron molecules, compact multiquarks, hybrid mesons and so on~\cite{2016-Chen-p1-121,2017-Ali-p123-198,2017-Lebed-p143-194,2018-Guo-p15004-15004,2019-Liu-p237-320,2020-Brambilla-p1-154}. Among these theoretical models, the loosely bound hadron molecule and compact multiquark are two especially appealing configurations. For the charmoniumlike XYZ and $P_c$ states, it is complicated and difficult 
to distinguish these two different hadron configurations experimentally and theoretically since the existence of light quarks. 

In 2017, an exotic structure around $18.4$ GeV was reported by CMS Collaboration in the $\Upsilon(1S)\mu^+\mu^-$ channel~\cite{CMS:2016liw}, which was once regarded as a fully-bottom $bb\bar b\bar b$ tetraquark state. In 2019, the ANDY Collaboration at RHIC reported an evidence of a significance peak at around 
$18.12$ GeV~\cite{ANDY:2019bfn}. Although these structures were not confirmed by some other experiments~\cite{LHCb:2018uwm,CMS:2020qwa}, 
their observations still attracted a lot of research interests in fully heavy tetraquark states~\cite{Chen:2016jxd,Anwar:2017toa,Esposito:2018cwh,Hughes:2017xie,Karliner:2016zzc,Wu:2016vtq,Richard:2017vry,Bai:2016int,Chen:2019dvd,Debastiani:2017msn}. Very recently, the LHCb Collaboration declared a narrow resonance $X(6900)$ in the di-$J\psi$ mass spectrum with the significance more than $5\sigma$~\cite{LHCb:2020bwg}. Moreover, a broad structure ranging from 6.2 to 6.8 GeV and a hint for another structure around 7.2 GeV were also reported at the same time~\cite{LHCb:2020bwg}. These exotic structures observed in LHCb immediately attracted great attention to study the fully-charm $cc\bar c\bar c$ tetraquarks for their mass spectra~\cite{Albuquerque:2020hio,Giron:2020wpx,Gordillo:2020sgc,Guo:2020pvt,Jin:2020jfc,Karliner:2020dta,Ke:2021iyh,Li:2021ygk,Liang:2021fzr,Liu:2019zuc,liu:2020eha,Pal:2021gkr,Sonnenschein:2020nwn,Wan:2020fsk,Wang:2018poa,Wang:2019rdo,Wang:2020ols,Wang:2021kfv,Weng:2020jao,Yang:2020rih,Yang:2020wkh,Zhang:2020xtb,Zhao:2020zjh,Zhao:2020nwy,Zhu:2020xni,Cao:2020gul,Mutuk:2021hmi,Yang:2021hrb}, their production mechanisms~\cite{Huang:2021vtb,Feng:2020riv,Wang:2020gmd,Feng:2020qee,Maciula:2020wri,Goncalves:2021ytq,Ma:2020kwb,Wang:2020tpt,Zhao:2020nwy,Zhu:2020sn,Gong:2020bmg} and their decay properties~\cite{Guo:2020pvt,Lu:2020cns,Li:2019uch,Chen:2020xwe,Becchi:2020uvq,Sonnenschein:2020nwn}. 
Since the absence of light quarks, a fully-heavy tetraquark system is more likely to form a compact tetraquark state via the gluon-exchange color interaction, but rather than a loosely hadron molecule combined by the light meson exchanged interaction~\cite{Maiani:2020pur,Chao:2020dml}. 

Nevertheless, the authors of Ref.~\cite{Dong:2021lkh} discussed the interaction between two $J/\psi$ mesons via the exchange of soft gluons, which hadronise into two light mesons at large distance. By studying the correlated $\pi\pi$ and $K\bar K$ exchanges, they found that it is possible for two $J/\psi$ mesons to form a bound state. 
In Ref.~\cite{Albuquerque:2020hio}, the authors studied the di-charmonia states in $\mathbf{1}_{[c\bar{c}]}\otimes \mathbf{1}_{[c\bar{c}]}$ configuration with $J^{PC}=0^{++}$ and predicted their masses around 6.0-6.7 GeV in the method of QCD sum rules. The existence of di-charmonia bound states were also studied in Ref.~\cite{Yang:2021hrb}, in which the authors 
investigated the $\eta_c\eta_c, J/\psi J/\psi$ bound states in both $\mathbf{1}_{[c\bar{c}]}\otimes \mathbf{1}_{[c\bar{c}]}$ and $\mathbf{8}_{[c\bar{c}]}\otimes \mathbf{8}_{[c\bar{c}]}$ color structures by using a non-relativistic  
quark model. 
The $0^{++}$ di-charmonia states in the $\mathbf{8}_{[c\bar{c}]}\otimes \mathbf{8}_{[c\bar{c}]}$ color structure were also investigated in Ref.~\cite{Yang:2020wkh} by the Laplace QCD sum rule method. 

In our previous works in Refs.~\cite{Chen:2016jxd,Yang:2021zrc,Wang:2021taf}, we have studied the fully-heavy tetraquark states in diquark-antidiquark 
configuration with both $\mathbf{6}_{[QQ]}\otimes \mathbf{\bar 6}_{[\bar{Q}\bar{Q}]}$ and $\mathbf{\bar 3}_{[QQ]}\otimes \mathbf{3}_{[\bar{Q}\bar{Q}]}$ color structures. In this work, we shall further investigate the possibility of fully-heavy tetraquark states in meson-meson configuration with $\mathbf{8}_{[Q\bar{Q}]}\otimes \mathbf{8}_{[Q\bar{Q}]}$ color structure by using the method of QCD moment sum rules~\cite{Reinders:1984sr,Shifman:1978bx}.

This paper is organized as follows. In Sec.~\ref{Sec_Current}, we construct the interpolating currents for $\mathbf{8}_{[Q\bar{Q}]}\otimes \mathbf{8}_{[Q\bar{Q}]}$ meson-meson tetraquark states. In Sec.~\ref{Sec_QCDSR}, we evaluate the correlation functions for these interpolating currents. We extract the masses for these tetraquark states by performing the moment QCD sum rule analyses in Sec.~\ref{Sec_Num}. The last section is a brief summary.

\section{Interpolating currents }\label{Sec_Current}
The color structure of a meson-meson operator $[Q \bar{Q}][Q \bar{Q}]$ can be written via the SU(3) symmetry
\begin{equation}
\begin{aligned}
(\mathbf{3} \otimes \overline{\mathbf{3}})_{[Q \bar{Q}]} \otimes(\mathbf{3} \otimes \overline{\mathbf{3}})_{[Q \bar{Q}]} &=(\mathbf{1} \oplus \mathbf{8})_{[Q \bar{Q}]} \otimes(\mathbf{1} \oplus \mathbf{8})_{[Q \bar{Q}]} \\
&=(\mathbf{1} \otimes \mathbf{1}) \oplus(\mathbf{1} \otimes \mathbf{8}) \oplus(\mathbf{8} \otimes \mathbf{1}) \oplus(\mathbf{8} \otimes \mathbf{8}) \\
&=\mathbf{1} \oplus \mathbf{8} \oplus \mathbf{8} \oplus(\mathbf{1} \oplus \mathbf{8} \oplus \mathbf{8} \oplus \mathbf{1 0} \oplus \overline{\mathbf{1 0}} \oplus \mathbf{27})\,,
\end{aligned}
\end{equation}
in which the color singlet structures come from the $\left(\mathbf{1}_{[Q \bar{Q}]} \otimes \mathbf{1}_{[Q \bar{Q}]}\right)$ and $\left(\mathbf{8}_{[Q \bar{Q}]} \otimes \mathbf{8}_{[Q \bar{Q}]}\right)$ terms. Following Ref.~\cite{Chen:2015ata}, we can construct the S-wave and P-wave $[Q \bar{Q}][Q \bar{Q}]$ interpolating currents as below: 
\begin{itemize}
\item The S-wave interpolating currents are
\begin{equation}
\begin{aligned}
J^{PC}=0^{++}:~&J_{1}=(\bar{Q}_{a}\gamma_{5}\lambda_{ab}^{n}Q_{b})  (\bar{Q}_{d}\gamma_{5}\lambda_{de}^{n}Q_{e})\, ,\\
       &J_{2}=(\bar{Q}_{a}\gamma_{\mu}\lambda_{ab}^{n}Q_{b})  (\bar{Q}_{d}\gamma_{\mu}\lambda_{de}^{n}Q_{e})\, ,\\
J^{PC}=1^{+-}:~&J_{1\mu}=(\bar{Q}_{a}\gamma_{\mu}\lambda_{ab}^{n}Q_{b})   (\bar{Q}_{d}\gamma_{5}\lambda_{de}^{n}Q_{e}) \, ,\\
J^{PC}=2^{++}, \, 0^{++}:~&J_{\mu\nu}=(\bar{Q}_{a}\gamma_{\mu}\lambda_{ab}^{n}Q_{b})  (\bar{Q}_{d}\gamma_{\nu}\lambda_{de}^{n}Q_{e})\, ,
\end{aligned}
\label{Jcurrents}
\end{equation}
in which we only obtain $[Q \bar{Q}][Q \bar{Q}]$ currents with $J^{PC}=0^{++}, 1^{+-}$ and $2^{++}$ in the S-wave channel. The tensor 
current $J_{\mu\nu}$ can couple to both the $J^{PC}=0^{++}$ and $2^{++}$ quantum numbers, but not the $J^{PC}=1^{++}$ channel since the Lorentz symmetry restriction. 

\item The P-wave interpolating currents are 
\begin{equation}
\begin{aligned}
J^{PC}=0^{-+}:~&\eta_{1}=(\bar{Q}_{a}\gamma_{5}\lambda_{ab}^{n}Q_{b})   (\bar{Q}_{d}\lambda_{de}^{n}Q_{e})\, ,\\
        &\eta_{2}=(\bar{Q}_{a}\sigma_{\mu\nu}\lambda_{ab}^{n}Q_{b})   (\bar{Q}_{d}\sigma_{\mu\nu}\gamma_{5}\lambda_{de}^{n}Q_{e})\, ,\\
J^{PC}=0^{--}:~&\eta_{3}=(\bar{Q}_{a}\gamma_{\mu}\lambda_{ab}^{n}Q_{b}) (\bar{Q}_{d}\gamma_{\mu}\gamma_{5}\lambda_{de}^{n}Q_{e})\, ,\\
J^{PC}=1^{--}:~&\eta_{1\mu}=(\bar{Q}_{a}\gamma_{\mu}\lambda_{ab}^{n}Q_{b})  (\bar{Q}_{d}\lambda_{de}^{n}Q_{e}) \, ,\\
       &\eta_{2\mu}=(\bar{Q}_{a}\gamma_{\alpha}\gamma_{5}\lambda_{ab}^{n}Q_{b})  (\bar{Q}_{d}\sigma_{\alpha\mu}\gamma_{5}\lambda_{de}^{n}Q_{e}) \, ,\\
J^{PC}=1^{-+}:~&\eta_{3\mu}=(\bar{Q}_{a}\gamma_{5}\lambda_{ab}^{n}Q_{b})  (\bar{Q}_{d}\gamma_{\mu}\gamma_{5}\lambda_{de}^{n}Q_{e})\, ,\\
       &\eta_{4\mu}=(\bar{Q}_{a}\gamma_{\mu}\lambda_{ab}^{n}Q_{b})  (\bar{Q}_{d}\sigma_{\mu\nu}\lambda_{de}^{n}Q_{e})\, ,
\end{aligned}
\label{etacurrents}
\end{equation}

\end{itemize}
where only one P-wave $[Q \bar{Q}]$ operator is contained in these interpolating currents. 
One should note that these tetraquark interpolating currents with $\mathbf{8}_{[Q\bar{Q}]}\otimes \mathbf{8}_{[Q\bar{Q}]}$ color structure can be written as combinations of the diquark-antidiquark operators through the Fierz transformation and the color rearrangement. Their decay properties should be the same with the $[QQ][\bar Q\bar Q]$ tetraquark states as discussed in Refs.~\cite{Chen:2016jxd,Chen:2020xwe}. Thus we shall not discuss the decay behaviors for these $[Q \bar{Q}][Q \bar{Q}]$ tetraquarks in this work. 

\section{QCD sum rules}\label{Sec_QCDSR}
In this section, we investigate the two-point correlation functions of the interpolating currents constructed above. For the scalar and pseudo-scalar currents, the correlation function are
\begin{equation}
\begin{aligned}
\Pi\left(p^{2}\right)&=i \int d^{4} x e^{i p \cdot x}\left\langle 0\left|T\left[J(x) J^{\dagger}(0)\right]\right| 0\right\rangle\, ,
\end{aligned}
\end{equation}
while for the vector and axial-vector currents
\begin{equation}
\begin{aligned}
 \Pi_{\mu \nu}\left(p^{2}\right) =i \int d^{4} x e^{i p \cdot x}\left\langle 0\left|T\left[J_{\mu}(x) J_{\nu}^{\dagger}(0)\right]\right| 0\right\rangle\, . 
 \label{CF_AV}
\end{aligned}
\end{equation}
The correlation function $\Pi_{\mu\nu} (p^{2})$ in Eq.~\eqref{CF_AV} can be divided into two part 
\begin{equation}
\Pi_{\mu \nu}\left(p^{2}\right)=\left(\frac{p_{\mu} p_{\nu}}{p^{2}}-g_{\mu \nu}\right) \Pi_{1}\left(p^{2}\right)+\frac{p_{\mu} p_{\nu}}{p^{2}}\Pi_{0}\left(p^{2}\right)\, ,
\end{equation}
where $\Pi_{0}\left(p^{2}\right)$ and $\Pi_{1}\left(p^{2}\right)$ represent the spin-0 and spin-1 invariant functions, respectively. 
For the tensor currents $J_{\mu\nu}(x)$ in Eq.~\eqref{Jcurrents}, 
\begin{equation}
\begin{aligned}
 \Pi_{\mu \nu,\rho \sigma}\left(p^{2}\right) =i \int d^{4} x e^{i p \cdot x}\left\langle 0\left|T\left[J_{\mu\nu}(x) J_{\rho\sigma}^{\dagger}(0)\right]\right| 0\right\rangle\, . 
 \label{CF_T}
\end{aligned}
\end{equation}
The correlation function $\Pi_{\mu\nu,\rho\sigma} (p^{2})$ in Eq.~\eqref{CF_T} can be expressed as 
\begin{equation}
\Pi_{\mu \nu,\rho\sigma} \left(p^{2}\right)=\left(\eta_{\mu\rho}\eta_{\nu\sigma}+\eta_{\mu\sigma}\eta_{\nu\rho}-\frac{2}{3}\eta_{\mu\nu}\eta_{\rho\sigma}\right) \Pi_{2}\left(p^{2}\right)+\cdots \, ,
\end{equation}
where
\begin{equation}\
\eta_{\mu\nu}=\frac{p_{\mu} p_{\nu}}{p^{2}}-g_{\mu \nu}\, .
\end{equation} 
The invariant function $\Pi_{2}\left(p^{2}\right)$ relates to the spin-2 intermediate state, and the $``\cdots"$ represents the contribution from the spin-0 
state.

 At the hadron level, the invariant functions can be expressed through the dispersion relation
\begin{equation}
\Pi\left(p^{2}\right)=\frac{\left(p^{2}\right)^{N}}{\pi} \int_{16m_{Q}^{2}}^{\infty} \frac{\operatorname{Im} \Pi(s)}{s^{N}\left(s-p^{2}-i \epsilon\right)} d s+\sum_{n=0}^{N-1} b_{n}\left(p^{2}\right)^{n}\, ,
\label{Cor-Spe}
\end{equation}
where $b_n$ is the subtraction constant. In QCD sum rules, the imaginary part of the correlation functions are usually simplified as the following “pole plus continuum” spectral function 
\begin{equation}
\rho (s)=\frac{1}{\pi} \text{Im}\Pi(s)=f_{H}^{2}\delta(s-m_{H}^{2})+\text{QCD continuum and higher states}\, ,
\end{equation}
in which the $\delta$ function represents the lowest-lying state. The parameters $f_{H}$ and $m_{H}$ are the coupling constant and mass of the lowest-lying hadronic resonance $H$ respectively 
\begin{equation}
\begin{aligned}
\langle 0|J| H\rangle &= f_{H}\, , \\
\left\langle 0\left|J_{\mu}\right| H\right\rangle &= f_{H} \epsilon_{\mu}\, , \\
\left\langle 0\left|J_{\mu\nu}\right| H\right\rangle &= f_{H} \epsilon_{\mu\nu}\, ,
 \end{aligned}
\end{equation}
with the polarization vector $\epsilon_{\mu}$ and polarization tensor $\epsilon_{\mu\nu}$.

To extract the lowest-lying resonance, we first define the moments by taking derivatives of the correlation function $\Pi(q^{2})$ in Euclidean region $Q^{2}=-q^{2}>0$
\begin{equation}
M_{n}\left(Q_{0}^{2}\right)=\left.\frac{1}{n !}\left(-\frac{d}{d Q^{2}}\right)^{n} \Pi\left(Q^{2}\right)\right|_{Q^{2}=Q_{0}^{2}}=\int_{16 m_{Q}^{2}}^{\infty} \frac{\rho(s)}{\left(s+Q_{0}^{2}\right)^{n+1}} d s\, .
\end{equation}
We then rewrite the moments by applying the above equation to the Eq.(\ref{Cor-Spe}) and obtain
\begin{equation}
M_{n}\left(Q_{0}^{2}\right)=\frac{f_{H}^{2}}{\left(m_{H}^{2}+Q_{0}^{2}\right)^{n+1}}\left[1+\delta_{n}\left(Q_{0}^{2}\right)\right]\,,
\end{equation}
where $\delta_{n}(Q_{0}^{2})$ represents  higher excited states and continuum contribution, and it is a function of $n$ and $Q_{0}^{2}$. It should be noted that for a specific value of $Q_{0}^{2}$, $\delta_{n}(Q_{0}^{2})$ will tend to zero as $n$ going to infinity. Considering the ratio of the moments to remove the unknown coupling constant $f_{H}$
\begin{equation}
r\left(n, Q_{0}^{2}\right) \equiv \frac{M_{n}\left(Q_{0}^{2}\right)}{M_{n+1}\left(Q_{0}^{2}\right)}=\left(m_{H}^{2}+Q_{0}^{2}\right) \frac{1+\delta_{n}\left(Q_{0}^{2}\right)}{1+\delta_{n+1}\left(Q_{0}^{2}\right)}\, ,
\end{equation}
where the relation $\delta_{n}(Q_{0}^{2})\approx \delta_{n+1}(Q_{0}^{2})$ will be satisfied when $n$ is large enough, and we can extract the hadron mass as 
\begin{equation}
m_{H}=\sqrt{r\left(n, Q_{0}^{2}\right)-Q_{0}^{2}}\, . \label{hadronmass}
\end{equation}

At the quark-gluon level, we can evaluate the invariant functions $\Pi(p^{2})$ via the operator product expansion (OPE) method. The Wilson coefficients can be calculated by adopting the following heavy quark propagator in momentum space
\begin{equation}
 i S_{Q}^{a b}(p)=\frac{i \delta^{a b}}{\hat{p}-m_{Q}}
 +\frac{i}{4} g_{s} \frac{\lambda_{a b}^{n}}{2} G_{\mu \nu}^{n} \frac{\sigma^{\mu \nu}\left(\hat{p}+m_{Q}\right)+\left(\hat{p}+m_{Q}\right) \sigma^{\mu \nu}}{12}
 +\frac{i \delta^{a b}}{12}\left\langle g_{s}^{2} G G\right\rangle m_{Q} \frac{p^{2}+m_{Q} \hat{p}}{(p^{2}-m_{Q}^{2})^{4}}\, , 
\end{equation}
where $Q$ represents the charm quark or bottom quark field. The superscripts $a, b$ are the color indices and $\hat{p}=p^{\mu}\gamma_{\mu}$. In this work, we will only evaluate the perturbative term and gluon condensate term in the correlation function, the contributions from higher non-perturbative terms such as tri-gluon condensate are small enough to be neglected.

\section{Numerical analysis}\label{Sec_Num}
We now perform the QCD moment sum rule analyses by adopting the following values of heavy quark masses and gluon condensate \cite{Nielsen:2009uh,Narison:2018nbv,ParticleDataGroup:2020ssz}
\begin{equation}
\begin{array}{l}
{m_{c}\left(m_{c}\right)=(1.27 _{-0.02}^{+0.02}) \mathrm{GeV}}\, , \vspace{1ex} \\
{m_{b}\left(m_{b}\right)=(4.18 _{-0.02}^{+0.03}) \mathrm{GeV}}\, , \vspace{1ex} \\
{\left\langle g_{s}^{2} G G\right\rangle= (0.88\pm0.25) \mathrm{GeV}^{4}}\, .
\end{array}
\end{equation}

\begin{figure}[h!]
\centering
\includegraphics[width=8cm]{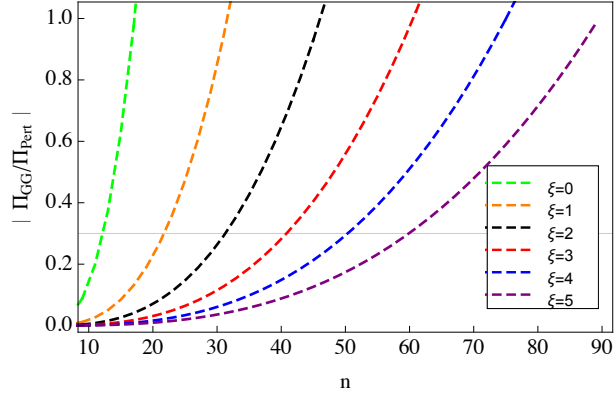}\\
\caption{$|\Pi^{GG}/\Pi^{Pert}|$ with respect to $n$ for different values of $\xi$ from $J_{1}(x)$ with $J^{PC}=0^{++}$.}
\label{GG-Pert-c}
\end{figure}

As mentioned in Sec.~\ref{Sec_QCDSR}, there remain two important parameters $n$ and $Q_{0}^{2}$ in the extracted hadron mass in Eq.~\eqref{hadronmass}. 
In the original literatures on moment sum rules, the authors set $Q_{0}^{2}=0$, which may lead to a bad OPE convergence. To avoid such bad behavior, we follow the Refs.~\cite{Chen:2016jxd,Yang:2021zrc,Wang:2021taf} to choose $Q_{0}^{2}>0$ and introduce $\xi=Q_{0}^{2}/(4m_{c})^{2}$ to perform sum rule analysis. The parameters $n$ and $\xi$ are related to each other through the following respects: (1) a large enough $n$ will reduce the higher excited states and continuum region contributions, but it will also decrease the convergence of OPE series. (2) a large $\xi$ (or $Q_{0}^{2}$) can compensate the OPE convergence (see Fig.~\ref{GG-Pert-c}), but may cause a bad convergence of $\delta_{n}(Q_{0}^{2})$ and make it difficult to obtain the parameters of the lowest lying resonance. One needs to find suitable working regions for these  two parameters to establish stable sum rules. 

We take the interpolating current $J_{1}(x)$ with $J^{PC}=0^{++}$ as an example to show the numerical analysis details. The correlation function of this current is evaluated as the following

\begin{align} 
\Pi^{pert}(Q^{2})&= \frac{1}{192\pi^{6}}\int_{0}^{1}dx\int_{0}^{1}dy\int_{0}^{1}dz
\Bigg\{\left( \frac{-13x(1-x)y(1-y)^{3}(1-z)}{z^{3}}\right)F(m_{c},Q^{2})^{4} +\left(\frac{40 m_{c}^{2}y(1-y)}{z^{2}}\right.  \nonumber\\
&\left. -\frac{52 Q^{2}x(1-x)y(1-y)^{3}(1-z)^{2}}{z^{2}}\right)F(m_{c},Q^{2})^{3}
+\left(\frac{60 m_{c}^{2}Q^{2}y(1-y)(1-z)}{z} -\frac{26 m_{c}^{4}(1-y)}{z^{2}}\right.   \nonumber\\
&\left.-\frac{26 Q^{4}x(1-x)y(1-y)^{3}(1-z)^{3}}{z}\right)F(m_{c},Q^{2})^{2}\Bigg\}\mbox{Log}\left[F(m_{c},Q^{2})\right]\, ,  
\nonumber\\
\Pi^{GG}(Q^{2})&= \frac{\langle g_{s}^{2}GG \rangle}{288\pi^{6}}\int_{0}^{1}dx\int_{0}^{1}dy\int_{0}^{1}dz
\Bigg\{\left(\frac{30m_{c}^{2} x (1-x)  (1-y)^{3} (1-z)^{3}}{z^{3}}-\frac{13 m_{c}^{2} x (1-x) y (1-y)^{3} (1-z)^{4}}{z^{3}}\right)  \nonumber\\
&\times F(m_{c},Q^{2}) +\left(\frac{10 m_{c}^{4} x (1-x) y (1-y)^{3} (1-z)^{4}}{z^{3}}-\frac{13 m_{c}^{4}(1-y)(1-z)^{2}}{z^{2}}-\frac{10 \text{mc}^4 (1-y) y (1-z)^3}{z^2}\right.   \nonumber\\
&+\left.\frac{15 m_{c}^{2} Q^{2} (1-x) x (1-y)^{3} (1-z)^{4}}{z^{2}}-\frac{26 m_{c}^{2} Q^{2} (1-x) x (1-y)^{3} (1-z)^{5}}{z^{2}}\right)\Bigg\}\mbox{Log}\left[F(m_{c},Q^{2})\right]  \nonumber\\
&+\frac{\langle g_{s}^{2}GG \rangle}{864\pi^{6}}\int_{0}^{1}dx\int_{0}^{1}dy\int_{0}^{1}dz \frac{-1}{F(m_{c},Q^{2})}
\Bigg\{
\frac{13m_{c}^{6}  (1-y) (1-z)^{3}}{z^{2}}-\frac{15 m_{c}^{4}Q^{2}y (1-y) (1-z)^{4}}{z^{2}}   \nonumber\\
&-\frac{15 m_{c}^{2}Q^{2} x (1-x) (1-y)^{3} (1-z)^{5}}{z^{2}}+\frac{13 m_{c}^{2}Q^{4} x (1-x) y (1-y)^{3} (1-z)^{6}}{z}\Bigg\}   \nonumber\\
&+\frac{\langle g_{s}^{2}GG \rangle}{1152\pi^{6}}\int_{0}^{1}dx\int_{0}^{1}dy\int_{0}^{1}dz \Bigg\{\left(\frac{18 (1-x) x (1-y)^{3} (1-z)^{2}}{z^{2}}\right)F(m_{c},Q^{2})^{2} 
 +\left(\frac{-12m_{c}^{2} (1-y) (1-z)}{z}  \right. \nonumber\\
&\left.-\frac{14 m_{c}^{2} x (1-x)(1-y)^{3} (1-z)^{2}}{y z^{2}}-\frac{24 m_{c}^2 y (1-z)}{x z}+\frac{36 Q^{2} x (1-x)  (1-y)^3 (1-z)^{3}}{z}\right)F(m_{c},Q^{2})\nonumber\\
&+\left(\frac{12 m_{c}^4 (1-y) (1-z)}{ y z}-\frac{12 m_{c}^2 Q^2 y (1-z)^2}{x}-6 m_{c}^2 Q^2 (1-y) (1-z)^2
\right.\nonumber\\
&\left.-\frac{7 m_{c}^2 Q^2 (1-x) x (1-y)^3 (1-z)^3}{ y z}+ 6 Q^4 (1-x) x (1-y)^3 (1-z)^4\right)\Bigg\}\mbox{Log}\left[F(m_{c},Q^{2})\right]\, ,
\end{align}
where $F(m_{c},Q^{2})=m_{c}^{2}\left(1-z+\frac{ z}{y}+\frac{ z}{x(1-y)}+\frac{z}{(1-x)(1-y)}\right)+Q^{2} z(1-z)$. We shall not evaluate the dimension-6 tri-gluon condensate $\langle G^3\rangle$ $(\sim$$g_s^3)$ and dimension-8 $\langle G^4\rangle$ $(\sim$$g_s^4)$ condensate in the OPE series. The 
$\langle G^3\rangle$ term gives negligible contribution to the correlation functions even at $\xi=0$ for the charmonium system~\cite{Nikolaev:1981ff} and four-charm tetraquark system~\cite{Zhang:2020xtb}. For the dimension-8 $\langle G^4\rangle$ condensate, it was proven in the charmonium moment sum rules that this term was much larger suppressed comparing to the dimension-4 gluon condensate $\langle G^2\rangle$ at $\xi\neq0$, and thus can also be neglected for the mass sum rule analysis~\cite{Nikolaev:1982ra,Reinders:1984sr}. 

To obtain convergent OPE series, we require that the contribution of the gluon condensate be smaller than the perturbative term, and obtain the upper bound $n_{max}=47, 61, 77, 91$ for $\xi=2, 3, 4, 5$ respectively. We show the ratio $|\Pi^{GG}/\Pi^{Pert}|$ in Fig.~\ref{GG-Pert-c} to display the convergence of the OPE series with respect to $n$ and $\xi$, which indicates that the OPE convergence becomes better with increasing of $\xi$ and decreasing of $n$. 

\begin{figure}[t!]
\centering
\includegraphics[width=10cm]{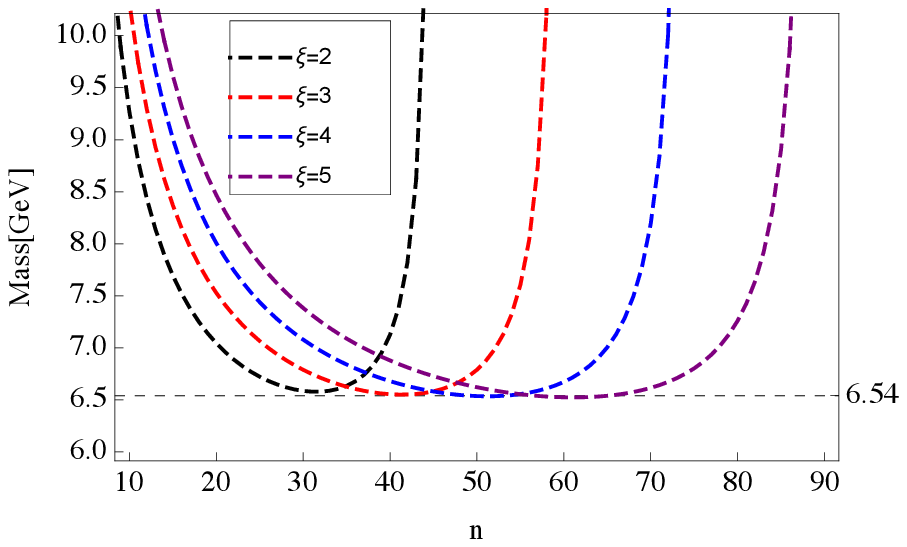}\\
\caption{Hadron mass for the fully-charm $c\bar cc\bar c$ tetraquark state with $J^{PC}=0^{++}$ from $J_{1}(x)$ with respect to $n$ for different value of $\xi$.}
\label{massVSn}
\end{figure}

In Fig.~\ref{massVSn}, we show the variation of the extracted mass with $n$ for different value of $\xi$, and get stable mass prediction plateaus $(n, \xi)=(32, 2), (42, 3), (52, 4), (62, 5)$. To choose the values of $\xi$, one should consider both the existence of the mass plateaus and the stability of the hadron mass for growing $\xi$. Both of these two criteria can be satisfied for $\xi\geq2$, as shown in Fig.~\ref{massVSn}. 
Accordingly, the mass of such a $c\bar cc\bar c$ tetraquark state is finally predicted to be
\begin{equation}
m_{{c\bar cc\bar c}}=6.54_{-0.18}^{+0.19} ~\text{GeV}\, ,
\end{equation}
in which the errors come from the uncertainties of $\xi$ and $n$, charm quark mass and the gluon condensate. 

The same numerical analyses can be done for the other interpolating currents in Eqs.~(\ref{Jcurrents}-\ref{etacurrents}). Then we obtain the masses for all $c\bar cc\bar c$ tetraquark states in $\mathbf{8}_{[c\bar{c}]}\otimes \mathbf{8}_{[c\bar{c}]}$ configuration in Table~\ref{ccccResultTab}. In this mass spectra, we predict three S-wave $c\bar cc\bar c$ tetraquarks with $J^{PC}=0^{++}, 1^{+-}$ and $2^{++}$ and four P-wave $c\bar cc\bar c$ tetraquarks with $J^{PC}=0^{-+}, 0^{--}, 1^{--}$ and $1^{-+}$. The masses are predicted to be around $6.3-6.5$ GeV for the S-wave states while $7.0-7.2$ GeV for the P-wave states.   Comparing to the mass spectra obtained 
in Ref.~\cite{Chen:2016jxd}, the masses for S-wave fully-charm tetraquark states are consistent with each other in both the $[c\bar c][c\bar c]$ and 
$[cc][\bar c\bar c]$ configurations. However, the P-wave $[c\bar c][c\bar c]$ tetraquarks are predicted to be 200-300 MeV higher than those in the diquark-antidiquark configuration~\cite{Chen:2016jxd}. 
In Table~\ref{ccccResultTab}, we also list the masses for some S-wave $c\bar cc\bar c$ tetraquark states in $\mathbf{8}_{[c\bar{c}]}\otimes \mathbf{8}_{[c\bar{c}]}$ configuration obtained by the Laplace QCD sum rule~\cite{Yang:2020wkh} and a non-relativistic quark model~\cite{Yang:2021hrb}. Our results for these $c\bar cc\bar c$ tetraquark states are in good agreement with those in Refs.~\cite{Yang:2020wkh,Yang:2021hrb}.

One notes that the two interpolating currents in the same channel ($J^{PC}=0^{++}, 0^{-+}, 1^{-+}, 1^{--}$) lead to almost the same hadron masses. To specify if these two currents couple to the same physical state or not, we calculate their cross correlation functions of two different currents with the same quantum number, e.g., the $J_{1}(x)$ and $J_{2}(x)$ with $J^{PC}=0^{++}$
\begin{equation}
\Pi_{12}\left(p^{2}\right)=i \int d^{4} x e^{i p \cdot x}\left\langle 0\left|T\left[J_{1}(x) J_{2}^{\dagger}(0)\right]\right| 0\right\rangle\, .
\end{equation}
Our calculations show that all these cross correlation functions are large enough and comparable to the diagonal correlators, implying that they couple to the same physical states. Since the two interpolating currents in the same channel give almost the same hadron masses, we don't reanalyze the mass sum rules by using their mixing current, avoiding more errors from the uncertain mixing angle.


\begin{table*}[t!]
\caption{The mass spectra for the fully-charm $c\bar cc\bar c$ tetraquark states in $\mathbf{8}_{[c\bar{c}]}\otimes \mathbf{8}_{[c\bar{c}]}$ color configuration.}
\renewcommand\arraystretch{1.3} 
\setlength{\tabcolsep}{2.em}{ 
\begin{tabular}{ccccc}
 \hline  \hline 
   Current         & $J^{PC} $        &Mass(\text{GeV)}   &Ref.~\cite{Yang:2020wkh}(\text{GeV)} &Ref.~\cite{Yang:2021hrb} (\text{MeV)}\\ \hline
   $J_{1}$         &    $0^{++}$       &    $6.54_{-0.18}^{+0.19} $   & $6.44_{-0.11}^{+0.11} $   & $6403 $  \vspace{1ex}  \\
    $J_{2}$        &    $0^{++}$      &    $6.36_{-0.16}^{+0.16} $    & $6.52_{-0.11}^{+0.11} $   & $6346 $  \vspace{1ex}  \\
    $J_{1\mu}$     &    $1^{+-}$      &    $6.47_{-0.17}^{+0.18} $    &     -                     & $6325 $  \vspace{1ex}  \\  
  $J_{\mu\nu}$    &   $2^{++}$        &    $6.52_{-0.17}^{+0.17} $    &     -                     & $6388 $  \vspace{1ex}  \\

   $\eta_{1}$      &    $0^{-+}$      &    $7.00 _{-0.20}^{+0.23} $    &    -                     &      -   \vspace{1ex}  \\
  $\eta_{2}$         &    $0^{-+}$   &    $7.02 _{-0.20}^{+0.24} $     &    -                     &      -   \vspace{1ex}  \\
  $\eta_{3}$       &    $0^{--}$     &    $7.00 _{-0.20}^{+0.23} $     &    -                     &      -   \vspace{1ex}  \\
   $\eta_{1\mu}$   &    $1^{--}$     &    $6.99_{-0.20}^{+0.23} $      &    -                     &      -   \vspace{1ex}  \\
   $\eta_{2\mu}$    &    $1^{--}$     &    $7.17_{-0.22}^{+0.28} $     &    -                     &      -   \vspace{1ex}  \\
    $\eta_{3\mu}$   &    $1^{-+}$     &    $6.98_{-0.19}^{+0.21} $     &    -                     &      -   \vspace{1ex}  \\
     $\eta_{4\mu}$  &    $1^{-+}$     &    $7.07_{-0.19}^{+0.21} $     &    -                     &      -   \vspace{1ex}  \\
     \hline\hline  
\label{ccccResultTab}
\end{tabular}
}
\end{table*}

\begin{figure}[t!]
\centering
\includegraphics[width=10cm]{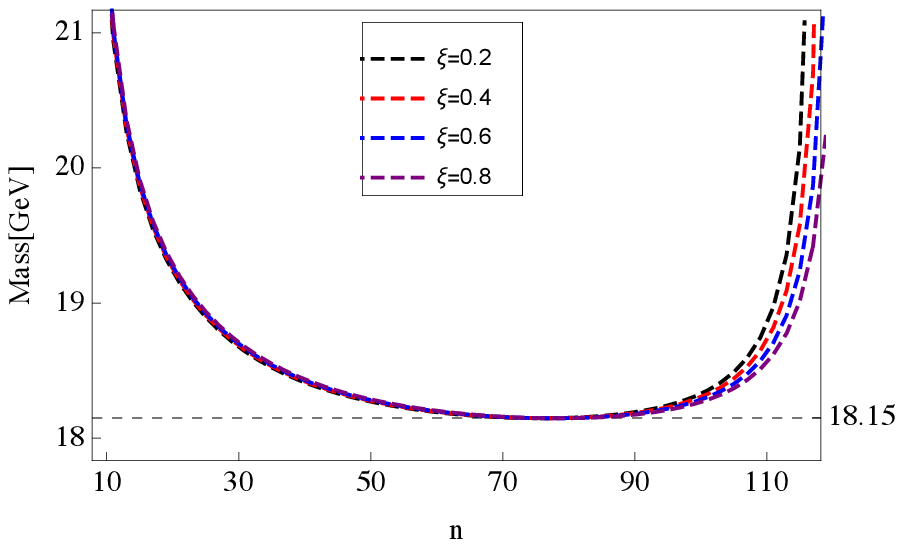}\\
\caption{Hadron mass for fully-bottom $b\bar bb\bar b$ tetraquark state  with $J^{PC}=0^{++}$ from $J_{1}(x)$ with respect to $n$ for different value of $\xi$.}
\label{massVSnb}
\end{figure}

We can also study the fully-bottom tetraquark states in $\mathbf{8}_{[b\bar{b}]}\otimes \mathbf{8}_{[b\bar{b}]}$ configuration. For the fully-bottom system, we define $\xi=Q_{0}^{2}/(m_{b})^{2}$ and find that the two criteria of mass plateaus and $\xi$ stability can be achieved for $\xi=0.2-0.8$. By requiring that the contribution of the gluon condensate be smaller than the perturbative term, we obtain the upper bound on the parameter $n_{max}=119, 121, 123, 125$ for $\xi=0.2, 0.4, 0.6, 0.8$ respectively. We show the variation of the extracted mass with $n$ for different value of $\xi$ in Fig.~(\ref{massVSnb}), and get stable mass prediction plateaus $(n, \xi)=(75, 0.2), (77, 0.4), (77, 0.6), (79, 0.8)$. The mass of such a $b\bar bb\bar b$ tetraquark state is finally predicted as
\begin{equation}
m_{b\bar bb\bar b}=18.15_{-0.10}^{+0.14} ~\text{GeV}\, .
\end{equation}
Applying the similar moment sum rule analyses, we obtain the mass spectra for these $b\bar bb\bar b$ tetraquark states and list them in Table~\ref{bbbbResultTab}. 
Accordingly, the S-wave $b\bar bb\bar b$ tetraquark states are obtained to be around 18.2 GeV while the P-wave states are about 18.4-18.8 GeV.  
Such results are several hundreds MeV below the masses of diquark-antidiquark $bb\bar b\bar b$ tetraquarks predicted in Ref.~\cite{Chen:2016jxd}. As shown in Table~\ref{bbbbResultTab}, our results for the S-wave $\mathbf{8}_{[b\bar{b}]}\otimes \mathbf{8}_{[b\bar{b}]}$ tetraquarks are much smaller than those predicted 
in the non-relativistic quark model~\cite{Yang:2021hrb}, but in roughly agreement with the results in Laplace QCD sum rules~\cite{Yang:2020wkh}.

\begin{table*}[t!]
\caption{The mass spectra for the fully-bottom $b\bar bb\bar b$ tetraquark states in $\mathbf{8}_{[b\bar{b}]}\otimes \mathbf{8}_{[b\bar{b}]}$ color configuration.}
\renewcommand\arraystretch{1.3} 
\setlength{\tabcolsep}{2.em}{ 
\begin{tabular}{ccccc}
 \hline  \hline 
   Current         & $J^{PC} $        &Mass(\text{GeV)}   &Ref.~\cite{Yang:2020wkh}(\text{GeV)} &Ref.~\cite{Yang:2021hrb} (\text{MeV)}\\ \hline
    $J_{1}$      &    $0^{++}$          &    $18.15_{-0.10}^{+0.14} $    &    $18.38_{-0.11}^{+0.11} $   & $19243 $ \vspace{1ex}  \\
    $J_{2}$       &    $0^{++}$         &    $18.13_{-0.09}^{+0.13} $    &    $18.44_{-0.10}^{+0.10} $   & $19237 $ \vspace{1ex}  \\
    $J_{1\mu}$    &    $1^{+-}$         &    $18.14_{-0.09}^{+0.14} $    &      -                        & $19126 $ \vspace{1ex}  \\  
  $J_{\mu\nu}$    &   $2^{++}$          &    $18.15_{-0.09}^{+0.14} $    &      -                        & $19197 $ \vspace{1ex}  \\

   $\eta_{1}$     &    $0^{-+}$         &    $18.45 _{-0.11}^{+0.15} $   &    -                          & -        \vspace{1ex}  \\
   $\eta_{2}$      &    $0^{-+}$        &    $18.54 _{-0.12}^{+0.16} $   &    -                          & -        \vspace{1ex}  \\
  $\eta_{3}$       &    $0^{--}$        &    $18.47 _{-0.11}^{+0.15} $   &    -                          & -         \vspace{1ex}  \\
   $\eta_{1\mu}$   &    $1^{--}$        &    $18.46_{-0.11}^{+0.15} $    &    -                          & -         \vspace{1ex}  \\

   $\eta_{2\mu}$     &    $1^{--}$      &    $18.46_{-0.11}^{+0.15} $    &    -                          & -         \vspace{1ex}  \\
    $\eta_{3\mu}$    &    $1^{-+}$      &    $18.56_{-0.11}^{+0.16} $    &    -                          & -         \vspace{1ex}  \\
     $\eta_{4\mu}$   &    $1^{-+}$      &    $18.79_{-0.13}^{+0.18} $    &    -                          & -         \vspace{1ex}  \\
    \hline\hline  
\label{bbbbResultTab}
\end{tabular}
}
\end{table*}

\newpage

\section{Conclusion and Discussion}
We have studied the fully-heavy $Q\bar{Q}Q\bar{Q}$ tetraquark states in the $\mathbf{8}_{[Q\bar{Q}]}\otimes \mathbf{8}_{[Q\bar{Q}]}$ color structure by using the moment QCD sum rule method. We construct the S-wave and P-wave interpolating tetraquark currents with various quantum numbers and calculate their two-point correlation functions containing perturbative term and gluon condensate term. Choosing suitable parameter working regions, we have established stable moment sum rules for all interpolating currents and extracted the mass spectra for the fully-charm and fully-bottom tetraquark states. 

For the fully-charm $c\bar{c}c\bar{c}$ system, our results suggest that the S-wave tetraquark states with $J^{PC}=0^{++}, 1^{+-}, 2^{++}$ lie around $6.3-6.5$ GeV while the P-wave tetraquark states with $J^{PC}=0^{-+}, 0^{--}, 1^{--}, 1^{-+}$ are about $7.0-7.2$GeV. 
Especially, the masses for the $c\bar cc\bar c$ tetraquarks with $J^{PC}=0^{++}$ and $2^{++}$ are consistent with the broad structure observed by LHCb~\cite{LHCb:2020bwg}. The P-wave fully-charm tetraquarks with $J^{PC}=0^{-+}$ and $1^{-+}$ are predicted to be roughly in agreement with the mass of $X(6900)$ within errors. Such results suggest the possibility that there are some $\mathbf{8}_{[c\bar{c}]}\otimes \mathbf{8}_{[c\bar{c}]}$ components in LHCb's di-$J/\psi$ structures. More investigations are needed in both theoretical and experimental aspects to study the nature of these structures.

For the fully-bottom $b\bar{b}b\bar{b}$ system, the numerical results show that the S-wave tetraquark states are about $18.2$ GeV while the P-wave states are around $18.4-18.8$ GeV. All these fully-bottom $b\bar{b}b\bar{b}$ tetraquarks are predicted to below the $\eta_b\eta_b$ and $\Upsilon(1S)\Upsilon(1S)$ two-meson decay thresholds, indicating that these tetraquark states will be stable against the strong interaction. Such results are consistent with our previous prediction for the diquark-antidiquark $bb\bar b\bar b$ tetraquarks in Ref.~\cite{Chen:2016jxd}. More efforts are expected to search for 
such fully-bottom tetraquark states in the future experiments, such as LHCb, CMS and so on.

\section*{ACKNOWLEDGMENTS}
This work is supported by the National Natural Science Foundation of China under Grant No. 12175318 and the National Key R$\&$D Program of China under Contracts No. 2020YFA0406400.


\end{document}